\documentclass[structabstract]{aa}  
\usepackage{graphicx}
\usepackage{txfonts}

\usepackage{natbib}
\bibpunct[, ]{(}{)}{;}{a}{}{,}
\newcommand{\griz}{$g^\prime,\, r^\prime,\, i^\prime,\, z^\prime$}
\newcommand{\uv}{$uvw2,\,uvm2,\,uvw1,\,u,\,b,\,v$}
\newcommand{\JHK}{$J,\, H,\, K_s$~}
\newcommand{\gK}{$g^\prime,\, r^\prime,\, i^\prime,\, z^\prime,\, J,\,H,\,K_s$}

\begin{document}
\title{Photometric redshifts for GRB afterglows from GROND and \textit{Swift}/UVOT}



\author{T. Kr{\"u}hler\inst{1,2}
\and P. Schady\inst{1}
\and J. Greiner\inst{1}
\and P. Afonso\inst{1}
\and E. Bottacini\inst{1}
\and C. Clemens\inst{1}
\and R. Filgas\inst{1} 
\and S. Klose\inst{3}
\and T. S. Koch\inst{4}
\and A. K\"{u}pc\"{u}-Yolda\c{s} \inst{5}
\and S. R. Oates \inst{6}
\and F. Olivares E.\inst{1}
\and M. J. Page \inst{6}
\and S. McBreen\inst{7} 
\and M. Nardini\inst{1}
\and A. Nicuesa Guelbenzu\inst{3} 
\and A. Rau\inst{1}
\and P. W. A. Roming\inst{4,8}
\and A. Rossi\inst{3} 
\and A. Updike \inst{9}
\and A. Yolda\c{s} \inst{1}}

\institute{Max-Planck-Institut f\"{u}r extraterrestrische Physik, Giessenbachstra{\ss}e, 85748 Garching, Germany
            \email{kruehler@mpe.mpg.de}
            \and
            Universe Cluster, Technische Universit\"{a}t M\"{u}nchen, Boltzmannstra{\ss}e 2, 85748, Garching, Germany
            \and
           Th\"{u}ringer Landessternwarte Tautenburg, Sternwarte 5, 07778 Tautenburg, Germany
           \and
           Department of Astronomy \& Astrophysics, Pennsylvania State University, 525 Davey Laboratory, University Park, PA 16802, USA
           \and
           European Southern Observatory, 85748 Garching, Germany
           \and 
           Mullard Space Science Laboratory, University College London, Holmbury St Mary, Dorking, Surrey RH5 6NT, UK
           \and
           School of Physics, University College Dublin, Dublin 4, Ireland
           \and
           Space Science Department, Southwest Research Institute, 6220 Culebra Rd, San Antonio, TX 78238, USA
                      \and 
           Department of Physics and Astronomy, Clemson University, Clemson, SC 29634, USA    
}


 
\abstract
{}
{We present a framework to obtain photometric redshifts (photo-$z$s) for gamma-ray burst afterglows. Using multi-band photometry from GROND and \textit{Swift}/UVOT, photo-$z$s are derived for five GRBs for which spectroscopic redshifts are not available.}
{We use UV/optical/NIR data and synthetic photometry based on afterglow observations and theory to derive the photometric redshifts of GRBs and their accuracy. Taking into account the afterglow synchrotron emission properties, we investigate the application of photometry to derive redshifts in a theoretical range between $z\sim1$ to $z\sim12$.}
{The measurement of photo-$z$s for GRB afterglows provides a quick, robust and reliable determination of the distance scale to the burst, particularly in those cases where spectroscopic observations in the optical/NIR range cannot be obtained. Given a sufficiently bright and mildly reddened afterglow, the relative photo-$z$ accuracy $\eta = \Delta z/(1+z)$ is better than 10\% between $z=1.5$ and $z\sim7$ and better than 5\% between $z=2$ and $z=6$. We detail the approach on 5 sources without spectroscopic redshifts observed with UVOT on-board \textit{Swift} and/or GROND. The distance scale to those same afterglows is measured to be $z=4.31^{+0.14}_{-0.15}$ for GRB~080825B, $z=2.13^{+0.14}_{-0.20}$ for GRB 080906, $z=3.44^{+0.15}_{-0.32}$ for GRB~081228, $z=2.03^{+0.16}_{-0.14}$ for GRB 081230 and $z=1.28^{+0.16}_{-0.15}$ for GRB~090530.}
{Due to the exceptional luminosity and simple continuum spectrum of GRB afterglows, photometric redshifts can be obtained to an accuracy as good as $\eta \sim 0.03$ over a large redshift range including robust ($\eta \sim 0.1$) measurements in the ultra-high redshift regime ($z\gtrsim 7$). Combining the response from UVOT with ground-based observatories and in particular GROND operating in the optical/NIR wavelength regime, reliable photo-$z$s can be obtained from $z \sim 1.0$ out to $z\sim 10$, and possibly even at higher redshifts in some favorable cases, provided that these GRBs exist, are localized quickly, have sufficiently bright afterglows and are not heavily obscured.}

\titlerunning{Afterglow photo-$z$s}
\authorrunning{T. Kr{\"u}hler et al.}
\maketitle
%

\section{Introduction}

The measurement of cosmological redshifts using photometry in broad-band filters avails of prominent and characteristic features in the spectra of extragalactic objects, such as the strong 4000~\AA~Ca H/K break in early-type galaxies \citep[e.g.][]{1962IAUS...15..390B}, or the Lyman-breaks for highly redshifted objects \citep[e.g.][]{1992AJ....104..941S, 1996ApJ...462L..17S, 1996MNRAS.283.1388M}. The crude spectral coverage of broad-band photometry compared to even low-resolution spectroscopy limits both the accuracy and the application of photo-$z$ measurements to a redshift range where the data contain significant redshift signatures. However, with the advent of large-field, deep multi-band photometric and spectroscopic surveys as the SDSS \citep[e.g.][]{2000AJ....120.1579Y}, the Hubble Deep Fields \citep[e.g.][]{1996AJ....112.1335W}, COSMOS \citep[e.g.][]{2007ApJS..172....1S}, or GOODS \citep[e.g.][]{2004ApJ...600L..93G} two major limitations of photometric redshifts were overcome: limited photometric accuracy and wavelength coverage and a lack of training sets of spectroscopic redshifts to calibrate and validate the photometric redshift determinations convincingly. Robust and accurate photo-$z$s can now be obtained for a large number of objects efficiently and to a depth which is inaccessible to spectroscopy at even the largest telescopes \citep[e.g.][]{1997AJ....113....1S, 1998AJ....115.1418H, 2003AJ....125..580C, 2003ApJ...587L..79F, 2004ApJ...600L.167M, 2007ApJS..172...99C, 2006A&A...457..841I, 2009ApJ...690.1236I, 2009ApJ...690.1250S, 2010ApJ...709L.133B, 2010MNRAS.403..960M}. 

A further application for photometric redshifts in addition to multi-object and faint-source distance determination is in the field of gamma-ray burst (GRB) astronomy \citep[see e.g.][Cucchiara et al., in prep.]{2005A&A...443L...1T, 2006A&A...447..897J, 2006Natur.440..181H, 2008A&A...490.1047C, 2008A&A...491L..29R, 2009A&A...498...89G, 2009MNRAS.395..490O, 2009Natur.461.1254T, 2009Natur.461.1258S, 2009GCN..9215....1O, 2010arXiv1003.3885M}. Despite its extreme early luminosity, a GRB afterglow fades quickly. This enables an unambiguous identification but requires that afterglow spectroscopy \citep[e.g.][]{2005ApJ...634..501B, 2006A&A...447..897J, 2006Natur.440..184K, 2007ApJS..168..231P, 2009ApJS..185..526F} must be performed in a timely manner in the first few hours to days after the prompt $\gamma$-ray emission. Rapid multi-band observations of the afterglow with small to medium aperture telescopes provide photometric redshift measurements in cases where spectroscopic observations do not reveal emission/absorption lines, or are unfeasible, for example when the optical/NIR afterglow is either too faint, has not been identified to arcsec accuracy or the spectroscopic resource is unavailable. Photometric follow-up on timescales of several tens of seconds to minutes and the resulting photo-$z$ also provides the trigger for a finetuned setup for long-slit spectrographs with limited wavelength coverage \citep[e.g.][]{2009ApJ...693.1610G, 2010arXiv1004.3261R}. 

The wealth of photometric redshift codes \citep[e.g.][]{1995AJ....110.2655C, bol00, 2000ApJ...536..571B, 2001defi.conf...96B, 2002astro.ph..3445T, 2004PASP..116..345C, 2010ApJ...712..511C} can essentially be sub-divided into learning based methods and template fitting. The earlier requires a sufficiently large training set of photometry together with spectroscopic redshifts, which although limits the application (object types, different redshift/magnitude space), is however self-contained and makes no prior assumptions on the physical properties of the unknown objects. In contrast, template methods assume a spectral shape as obtained from observations and models of similar objects and compare its synthetic photometry with observations of the source of interest, which extrapolates reasonably well into unmeasured redshift ranges. The probability of a redshift solution is then evaluated using, for example, $\chi^2$ or Bayesian statistics.

The natural approach for GRB afterglows is template based, as it incorporates the well-known emission properties of the light source. The emission of GRB afterglows arises when the ultra-relativistic ejecta from the GRB central engine are decelerated by the swept-up circumburst medium \citep[e.g.][]{1993ApJ...418L...5P, 1997ApJ...476..232M, 2004RvMP...76.1143P, 2006RPPh...69.2259M, 2007ChJAA...7....1Z, 2009ARA&A..47..567G}. The optical afterglow is released via synchrotron emission from external shocks, where the kinetic particle energy is transformed into radiation \citep[e.g.][]{1997MNRAS.288L..51W}. Hence, the theoretical continuum spectrum of afterglows is a three-fold broken power law, with breaks at the self-absorption frequency $\nu_a$, the injection or typical frequency $\nu_i$ and the cooling frequency $\nu_c$ \citep{1998ApJ...497L..17S, 2000ApJ...543...66P, 2002ApJ...568..820G}. In the late afterglow, $\nu_i<\nu_c$ (the slow cooling regime) and the optical wavelength range is located either above or below $\nu_c$, which is fully supported by current observations \citep[e.g.][]{2001ApJ...549L.209G, 2004ApJ...608..846S, 2006A&A...451..821N, 2007MNRAS.377..273S, 2010AandAJochen}. Two prominent signatures, the Ly-$\alpha$ and Lyman-limit break at $121.5\times(1+z)$~nm and $91.2\times(1+z)$~nm respectively, and their characteristic redshift-dependent prominence, allow for a robust and precise redshift determination.

Here we present a framework of GRB afterglow photo-$z$ measurements, investigate their uncertainties, and detail the approach on several afterglows observed with the \textit{Swift}/UltraViolet Optical Telescope (UVOT; \citealp{2004ApJ...611.1005G, 2005SSRv..120...95R}) covering the 170-600~nm wavelength range in six filters and the Gamma-Ray burst Optical/NearInfrared Detector (GROND; \citealp{2007Msngr.130...12G, 2008PASP..120..405G}) which is sensitive between 360~nm and 2300~nm using seven filter bands (\gK).

Throughout the paper we adopt the convention that the flux density of the afterglow $F_\nu(\nu, t)$ can be described as $F_\nu(\nu, t) \propto \nu^{-\beta}t^{-\alpha}$, and concordance ($\Omega_M=0.27$, $\Omega_{\Lambda}=0.73$, $H_0=71$~km/s/Mpc) cosmology. All errors are given at 1$\sigma$ confidence unless indicated otherwise.

\section{Photometric Redshifts for GRB afterglows}
\label{photozgrb}

\subsection{Filters and Photometry}

\begin{figure}
\centering
\includegraphics[width=\columnwidth]{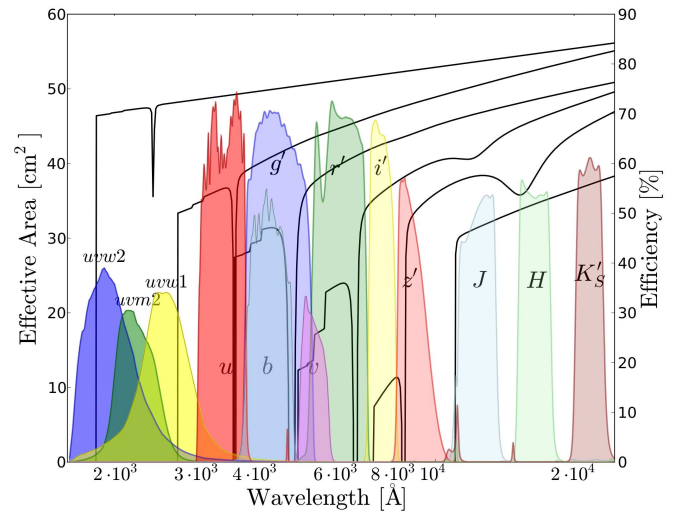}
\caption{UVOT (\uv) effective areas (against left y-axis) and GROND (\gK) filter curves (right y-axis), respectively in colored areas. The UVOT effective areas have been obtained from the most recent HEASARC Calibration Database$^1$. GROND filter curves include all optical components in GROND including the telescope, but exclude the atmosphere. Shown in black lines are template afterglow spectra, from redshifts $z$=1, 2, 3, 4.5, 6 to 8 (Top left to bottom right). These spectra also differ in their spectral index and rest-frame extinction (amount and reddening law), see e.g. the two synthetic spectra at $z$=4.5 and 6, which show a redshifted 2175\AA~dust feature. Also the DLAs centered at $121.5\times (1+z)$~nm differ in their hydrogen column densities.}
\label{filters}
\end{figure}

The key ingredient for any photometric redshift measurement is high-quality photometry over a large wavelength range. For GRB afterglows \textit{Swift}/UVOT and GROND offer the natural data source as both instruments systematically follow-up on GRB triggers and nicely complement each others sensitivity and wavelength coverage. UVOT onboard the \textit{Swift} satellite is a 30~cm space-based telescope primarily sensitive in the ultraviolet (UV) and optical range using \uv~ filters, that starts observing the GRB field as quickly as $\sim$40~s after the trigger \citep{2009ApJ...690..163R}. Additional optical and near-infrared (NIR) response is provided by GROND, a seven channel imager (\gK~simultaneously) mounted at the 2.2~m MPI/ESO telescope at the ESO/LaSilla observatory. GROND's reaction to GRB triggers is hence subject to visibility constraints, and is typically in the range of a couple to several hours post-burst. The effective area of the UVOT filters\footnote{http://heasarc.gsfc.nasa.gov/docs/heasarc/caldb/} and GROND's sensitivity curves are all shown in Fig.~\ref{filters}.

\subsection{General constraints on afterglow photo-$z$s}

The photometric identification of a redshift signature in the spectral energy distribution (SED) of a GRB afterglow requires detections in at least two filters to measure the continuum and another one to locate the Lyman-break. In this minimal case, the continuum is not well constrained, of course, and a strong degeneracy exists between afterglow spectral index and intrinsic reddening. Consequently, the redshift information is rather crude. Additional observations in independent wavelength ranges provide an improved continuum determination or constraints on the spectral index, and result in a more accurate photo-$z$ determination. The condition of having at least three individual filters might possibly be relaxed if further information  about the spectral index or the dust content of the afterglow is available from X-ray measurements as provided by \textit{Swift}/XRT \citep{bur05} and the optical to X-ray flux ratio \citep[e.g.][]{2010AandAPaulo}. In the standard model, for example, the optical spectral index $\beta_{\rm o}$ is required to be $\beta_{\rm o}=\beta_{\rm X}$ if both wavelength ranges probe the same part of the synchrotron spectrum, or $\beta_{\rm o}=\beta_{\rm X}-0.5$ in case of a cooling break lying in between \citep[e.g.][]{1998ApJ...497L..17S, 2002ApJ...568..820G}. Also, a large $A_V^{\rm host}$ solutions could be considered unlikely in cases where the combined fit requires the de-reddened optical to X-ray flux ratio $\beta_{\rm {oX}} > \beta_{\rm X}$. This is when the extrapolation of the dust-corrected UV/optical/NIR SED would significantly over-predict the X-ray measurement. Similar combined optical/X-ray SED analysis make however strong assumptions on the nature of the emission in X-ray and optical energy ranges, in particular that both are emitted by the same population of radiating electrons. The differences in the respective light curves \citep[e.g.][]{2006MNRAS.369.2059P}, and their inconsistencies with the most simple fireball scenarios \citep[e.g.][]{2007ApJ...662.1093W, 2009MNRAS.397.1177E} raises questions, whether this is indeed the case for all afterglows. Therefore, the most reliable approach is solely based on UV/optical/NIR data.

Even in the case of a well-sampled SED, several systematic uncertainties limit the accuracy of the photometric redshift determination, where the most dominant ones are the uncertainty in the dust-reddening properties, in the opacity of the Ly-$\alpha$ forest and the strength of the Damped Lyman-$\alpha$ Absorber (DLA) associated with the GRB. To quantify the systematic effects and the accuracy of the redshift measurement via photometry, a large sample of 4000 GRB afterglow SEDs was simulated as described in Sec.~\ref{synsample} and analyzed following Sec.~\ref{photozcode}.

\subsection{A synthetic afterglow spectral energy distribution}
\label{synsample}

The continuum emission of an afterglow in the UV/optical/NIR range is synchrotron radiation, and usually described as a single power law ($F_{\nu}(\lambda)=F_{0}(\lambda/\lambda_0)^{\beta}$, for caveats on this assumption, see Sec.~\ref{caveats}). This continuum spectrum is affected by reddening due to dust in the host galaxy. Therefore, different dust columns with average attenuation laws as observed in the Milky Way (MW), Small and Large Magellanic Clouds (SMC, LMC) in the parametrization according to \citet{1992ApJ...395..130P} between the burst and the observer are added to the synthetic spectrum. In addition, a more generic approach is used to obtain a broader range of plausible dust extinction laws via the Drude model proposed in \citet{li08} and \citet{2009ApJ...690L..56L}. Local dust-reddening laws strongly differ in their absolute UV absorption, which for a given $A_V$ decreases in strength from SMC, over LMC to MW. The most prominent extinction feature is the 2175~\AA~bump, which is generally attributed to absorption by graphite grains \citep[e.g.][]{1965ApJ...142.1681S, 2003ARA&A..41..241D}. While the feature is highly significant in MW and LMC models, it is absent in the SMC dust attenuation law. Although most bright afterglows are best described with featureless reddening similar as observed in the SMC \citep[e.g.][]{2004ApJ...608..846S, 2006ApJ...641..993K, 2007ApJ...661..787S, 2007MNRAS.377..273S, 2010MNRAS.401.2773S}, several recent observations show a strong 2175~\AA~bump in the afterglow's SED at the redshift of the burst \citep[e.g.][]{2008ApJ...685..376K, 2009ApJ...691L..27P, 2009ApJ...697.1725E, 2009arXiv0912.5435D, 2010arXiv1009.0004P}, where its location can be used as a redshift tracer in case the photometric data are well sampled.

A large number of optical afterglow spectra also show the presence of a strong DLA \citep[e.g.][]{2001A&A...370..909J, 2004A&A...419..927V, 2005ApJ...634L..25C, 2006ApJ...652.1011W, 2006NJPh....8..195S, 2006A&A...451L..47F, 2006ApJ...642..979B, 2006A&A...460L..13J, 2007ApJ...666..267P, 2009ApJS..185..526F}. To quantify the effects of neutral hydrogen absorption associated with the GRB host, DLAs with different hydrogen columns are added to the afterglow spectrum following the description of \citet{2006PASJ...58..485T} and references therein.

Bluewards of the DLA centered at the redshifted $\lambda_{\alpha}$, i.e. at $121.5(1+z)~\rm{nm}$ in the observers frame, the afterglow flux is further suppressed by the Ly-$\alpha$ forest \citep{1971ApJ...164L..73L}: intervening neutral hydrogen absorbers between the burst site and the observer. To account for different measurements, measurement errors and specific sight lines, the Ly-$\alpha$ effective optical depth $\tau_{\rm{Ly-\alpha}}$ is allowed to vary according to the constraints given in Table~4 of \citet{2008ApJ...681..831F} in the redshift range between 2 and 4.2. Below $z\sim 2$, $\tau_{\rm{Ly-\alpha}}$ is essentially zero, and above $z\sim4.2$, $\tau_{\rm{Ly-\alpha}}$ and its error are extrapolated from the lower-redshift data. The derived optical depth $\tau_{\rm{Ly-\alpha}}$ is then converted into an averaged flux depression factor $\langle D \rangle$ \citep{1982ApJ...255...11O, 1995ApJ...441...18M}.

Combining the upper effects, the synthetic afterglow spectrum becomes (see Fig.~\ref{filters} for examples):
\begin{equation}
F_{\nu}(\lambda, \vec{x})=\langle D_{\alpha}(z)\rangle F_{0} \left(\frac{\lambda}{\lambda_0}\right)^{\beta} \exp [-\tau_{\rm dust}(z, A_V)-\tau_{\rm DLA}(z, N_H)]
\end{equation}
for $\lambda_{\beta}(1+z) < \lambda_{\rm obs} < \lambda_{\alpha}(1+z)$ and similar for $\langle D_{\rm \beta}\rangle$ and higher order hydrogen absorptions. 
Here, $\vec{x}$ are the free parameters $\vec{x}=(z, A_V, \beta, N_H)$. Below the Lyman-limit at $\sim91.2(1+z)$~nm, we assume that the observed flux is fully attenuated by the neutral hydrogen along the line of sight, and hence $F_{\nu}(\lambda) = 0$. The intrinsic brightness term $F_0$ of the afterglows is selected to be in a range of previously observed UVOT/GROND afterglows and accounts for the typical reaction time of the instruments.

\begin{figure}
\centering
\includegraphics[width=\columnwidth]{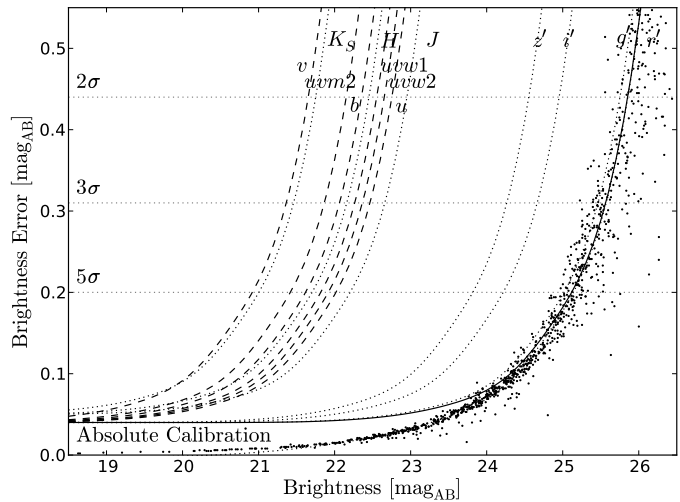}
\caption{Relation between brightness of an object and photometric accuracy detailed for a 2~h GROND and 600~s UVOT exposure under fairly typical observing conditions (3 days from new moon, airmass 1.5, 1~arcsec seeing). The black dots show the brightness and photon noise of the individual objects in the $r^\prime$-band, while the solid line denotes the relation between object brightness and total (photon+calibration) magnitude error. Dotted (for GROND) and dashed (for UVOT) lines show the same relation for the remaining filters. Individual objects are not plotted for these filters to enhance clarity. Horizontal dashed lines indicate typical limiting magnitudes for 2$\sigma$, 3$\sigma$ and 5$\sigma$ confidence levels.}
\label{magerr}
\end{figure}

Synthetic AB magnitudes of the afterglow in the different filters $i$ are derived via:

\begin{equation}
\rm{mag}_{\rm{AB}}^i = -2.5\log\frac{\int \lambda^{-1} F_{\nu}(\lambda, \vec{x})T^i(\lambda)d\lambda}{\int \lambda^{-1} T^i(\lambda)d\lambda} + 23.9~{\rm mag}
\end{equation}
where $T(\lambda)$ are the specific filter curves as shown in Fig.~\ref{filters}, and $F_{\nu}$ is given in $\mu$Jy. The synthetic afterglow magnitudes are then varied using a Gaussian probability distribution with the associated photometric errors as its standard deviation. These errors are the superposition of a constant, i.e. minimum error according to the absolute photometric accuracy obtained with standard calibration and a brightness dependent term related to photon statistics. The earlier is conservatively set to 0.04~mag for \uv~and \griz~and 0.05~mag for \JHK \citep[cp.][]{2008MNRAS.383..627P, 2008PASP..120..405G}. The latter is a function of the brightness of the simulated afterglow, and takes into account the sensitivities of UVOT and GROND as well as their reaction time to GRB triggers. In detail, the brightness of the simulated afterglow has been mapped to a magnitude error using the image statistics of standard follow-up observations (2~h GROND integration at 8~h after the trigger, 10~min UVOT integration at 1~h after the trigger) as shown in Fig.~\ref{magerr}.

The resulting magnitudes and associated errors are then used to derive a photo-$z$ as described in Section~\ref{photozcode} to compare in- and output values.

\subsection{photo-$z$ code}
\label{photozcode}

The software used to derive afterglow photometric redshifts is based on the publicly available \textit{hyperZ} \citep{bol00}, which minimizes the $\chi^2$ from synthetic photometry of a template spectrum $F_\nu(\lambda, \vec{x})$ against the observed data, i.e.:

\begin{equation}
\chi^2(\vec{x}) = \sum_{i}\left[\frac{F_{\nu, \rm obs}(\lambda_i)-\rm{const.}\times F_\nu(\lambda, \vec{x})}{\sigma_i}\right]^2
\end{equation}

where $F_{\nu, \rm obs}(\lambda_i)$ denotes the measurements in different broad band filters i with associated errors $\sigma_i$.

In addition to the existing treatment of the Lyman absorption according to \citet{1995ApJ...441...18M} and default reddening templates \citep{1976asqu.book.....A, 1979MNRAS.187P..73S, 1986AJ.....92.1068F, 1984A&A...132..389P, 1985A&A...149..330B,1989ApJ...345..245C,  2000ApJ...533..682C}, the code is complemented by several additions. In particular, power-law spectral templates with the possibility to constrain the spectral index, a DLA and a reddening law following \citet{mai04} were added. It also includes the measured total UVOT and GROND filter response, including all optical components in the pathway from the primary mirror of the telescope to the quantum efficiency of the detectors (see Fig.~\ref{filters} and also \citealp{2008PASP..120..405G, 2008MNRAS.383..627P}).

The results are similar to the standard outputs of the original \textit{hyperZ} version, including the photometric redshift, error bars or contours of the three main parameters (redshift, spectral index and intrinsic reddening) at arbitrary confidence levels and the statistical probability associated with the derived redshifts and secondary solutions when relevant.

\subsection{Properties of the afterglow mock sample}

\begin{figure}
\centering
\includegraphics[width=\columnwidth]{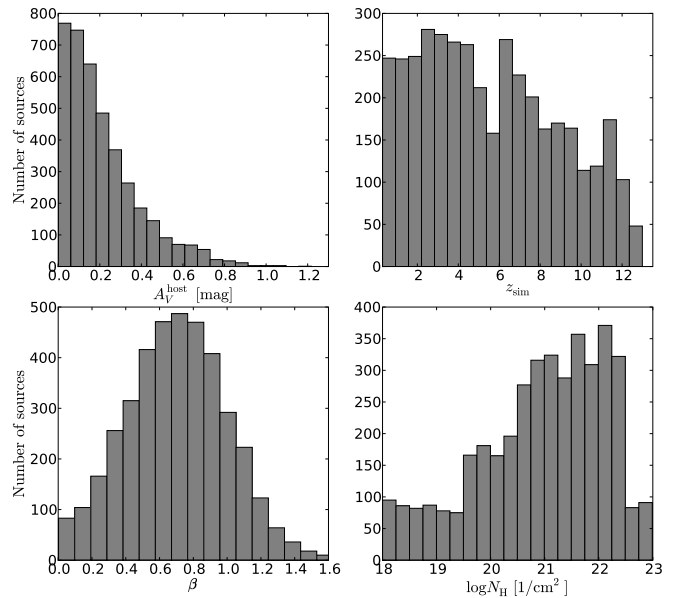}
\caption{Properties of the sample of 4000 simulated GRB afterglow spectra with respect to their host extinction $A_{V}^{\rm host}$, spectral index $\beta$, simulated redshift $z_{\rm sim}$, and DLA hydrogen column density $\log(N_{\rm H})$ from top left to bottom right.}
\label{sample}
\end{figure}

The properties of the mock set of 4000 afterglow spectra are shown in Fig.~\ref{sample}. The sample properties are chosen to be as close as possible to what is known about optical afterglows with respect to their spectral indices \citep[e.g.][]{2007arXiv0712.2186K} and the neutral hydrogen column densities of their DLAs \citep[e.g.][]{2009ApJS..185..526F} but are not fully representative of the global GRB afterglow properties. All previous demographic studies of GRB afterglows are still strongly biased against highly-reddened and redshifted bursts, and complete afterglow properties are hence subject to large uncertainties. Instead, the simulated sample is used for the determination of typical accuracies and systematic effects in the photo-$z$ measurement, when the afterglow is detected by UVOT and/or GROND. In particular, the dust distribution which is based on \citet{2007arXiv0712.2186K}, \citet{2007MNRAS.377..273S, 2010MNRAS.401.2773S} and \citet{2010AandAJochen} is probably heavily skewed towards low dust environments. It hence rather resembles the sight lines towards bright and mildly extinguished afterglows as typically detected in the UV/optical range \citep{2006ApJ...641..993K, 2007MNRAS.377..273S, 2009ApJ...690..163R}. The redshift distribution is chosen to peak at the redshift interval where most of the \textit{Swift} bursts originate from ($z\sim1-4$, \citealp{2009ApJS..185..526F}), but includes a significant number of ultra-high redshift GRBs ($\sim$1000 at $z>8$) to investigate the application of photometric redshifts at these extreme values.

\subsection{The application and accuracy of afterglow photo-$z$s}
\label{photozacc}
Figure~\ref{accuracy} shows a comparison between redshifts of previous afterglows obtained from spectroscopy against photo-$z$s derived from GROND and UVOT data (see also Tab.~\ref{tab:photoz}) and summarizes the result of the simulation. In the simulation, all available extinction laws (see Sec.~\ref{photozcode}) are used in the analysis, and the reported photo-$z$ is based on the reddening law that returns the best-fit SED, i.e. the minimum $\chi^2$.

Below $z\sim0.8$, even UV photometric measurements do not cover the wavelength range of the Lyman-limit absorption, and the redshift is hence unconstrained. The sensitivity towards low-redshift events begins at $z\sim1.0$, when the Lyman-limit cuts off a significant amount of flux in the $uvw2$ filter. The relative uncertainty $\eta=\Delta z/(1+z)$ in the low-redshift regime $z\lesssim1.5$, however, is still comparatively large with $\eta\gtrsim 0.08$. In addition, there is a significant fraction of afterglows ($\sim 20-40$\%), where the photometric data do not provide redshift constraints ($\Delta z \lesssim -0.5$). The observations do not cover the part of the spectrum blue-wards of the spectral break, and Ly-limit absorption is somewhat degenerate to intrinsic dust extinction for UV detections with low S/N and large associated photometric errors. The redshift signature of the Ly-break can hence not be reliably identified in these cases. This effect is clearly visible in Fig.~\ref{accuracy} as an under-prediction of the average redshift of the sample by photometric measurements in this redshift interval.

The dust-redshift degeneracy is broken with increasing redshift, as the Ly-limit moves to redder wavelengths, producing a drop-out in $uvw2$ at $z\gtrsim1.7$ and in $uvm2$ at $z\gtrsim2.0$, which is too sharp to be mimicked by dust. Accordingly, $\eta$ decreases to $\eta\sim 0.05$ over this redshift range. At higher redshifts, the absolute redshift uncertainty stays roughly constant and hence the relative scatter decreases to $\eta \sim 0.04$ at $z\sim 4$ until redshifts of $z\sim5$, when it further drops to $\eta \lesssim 0.03$. At these redshifts, absorption in the Lyman-$\alpha$ forest becomes significant (see Fig.~\ref{filters}), and hence there are two signatures in the spectrum which are used to measure the redshift. Furthermore, there are enough filters both blue-wards, and red-wards of the breaks to measure their location and the continuum with high accuracy. 

At even higher redshift, the accuracy remains essentially constant until $z\sim 6.5$. This demonstrates that the total numbers of individual filters does not strongly affect the robustness of the photo-$z$ measurement, as long as the continuum is fairly well determined. The number of filter bands that contain constraining information decreases from 13 at $z\sim2$ to 5 at $z\sim 6.5$. Due to the intrinsic power-law spectrum of GRB afterglows, just a few filters red-wards of the break are sufficient to reliably measure the continuum, at least in the case of blue ($\beta \lesssim 0.5$, $A_V\lesssim 0.2$ with 2 filters) or mildly red ($\beta \lesssim 1.2$, $A_V\lesssim 0.4$ with 3 filters) events. Information about the location of either Ly-$\alpha$ or the Ly-limit is then derived via 1 or 2 filters covering the wavelength range at or blue-wards of the break(s).

As the wavelength spacing between $z^\prime$ and $J$ is relatively large, the photometric redshift is rather loosely constrained between $z \sim 6.5$ and $z\sim8.3$, increasing from $\eta\sim 0.04$ or $\Delta z \sim 0.3$ at $z = 6.5$ to $\eta \sim 0.1$ or $\Delta z \sim 1$ at a redshift of around 8. Also at redshifts $z\gtrsim 8$, photometric colors can put strong constraints on the redshift. If the source is relatively bright as compared to the sensitivity limit (as input in the simulations), the $H-K_s$ color, coupled with the flux decrease in $J$ and the $z^\prime$-band upper limit, yields a robust redshift measurement ($\eta \sim 0.1$) until $z\sim9.5$. There are, however, a significant number of rather strong outliers with $\Delta z \gtrsim 1.5$ in this region, which are associated with moderate dust reddening along the line of sight. Above a redshift of $z\sim 10$, the transmitted flux in $J$ is below the instruments sensitivity limit even for intrinsically bright afterglows. With detections in only $H$ and $K_s$, the measurement is fully degenerate between the spectral power-law slope $\beta$, and in particular redshift and reddening. Only if the $J$-band limit is deep enough to exclude dust as the cause of the extremely red SED, a rough (around $\Delta z \sim 1-2$) redshift estimate might be obtained under specific assumptions and possibly in combination with X-ray observations of the fading afterglow until redshifts of around $z\sim 12$. 

The accuracy of photometric redshifts is further demonstrated and verified by photo-$z$s obtained for a number of afterglows where spectroscopic redshifts could be secured (red data in Fig.~\ref{accuracy}). These sources span a redshift range from $z=1.67$ to $z\sim8.2$, and support the results of the simulation.

\subsection{Systematic effects}
\begin{figure}
\includegraphics[width=\columnwidth]{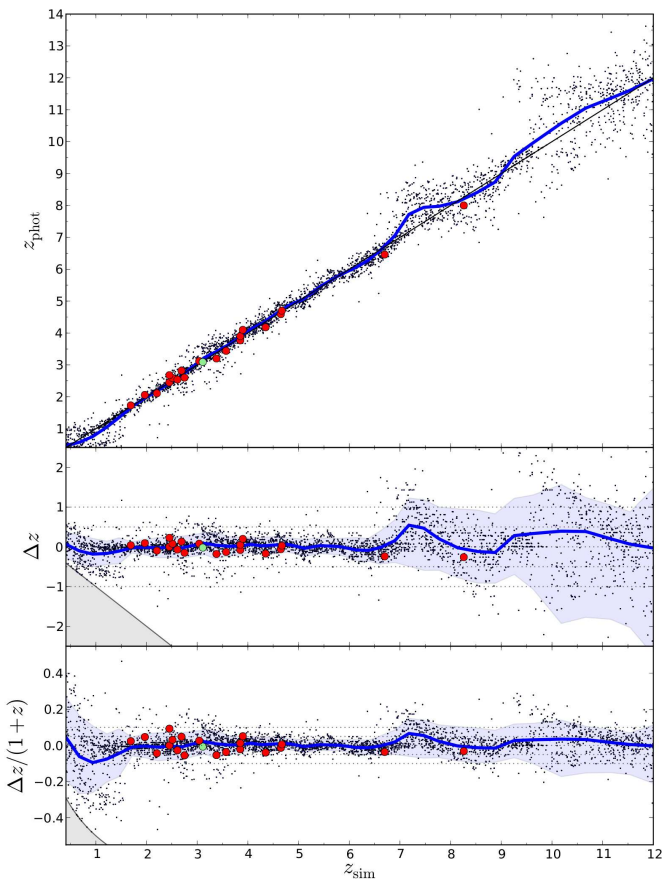}
\caption{Photometric redshift accuracy. Small black dots show the mock set of simulated afterglow spectra and their corresponding photo-$z$. Thick blue lines show the average photometric redshift after distributing the 4000 mock afterglows into redshift bins of 100 afterglows each. The lowest two panels also show in blue-shaded areas the quadratic sum of the typical difference to the input redshift and the 1$\sigma$ statistical uncertainty of the photo-$z$ analysis averaged over 100 afterglows in absolute ($\Delta z = z_{\rm phot}-z_{\rm sim}$) as well as relative ($\eta = \Delta z/(1+z)$) terms. The statistical probability that the true value is in the interval of +/-20\% around the quoted 1$\sigma$ accuracy is $>$95\%. Grey shaded areas represent the $z_{\rm phot}>0$ constraint. The thin, horizontal, dotted lines in the lowest two panels represent $\Delta z = $-1, -0.5, 0, +0.5, +1 and $\eta = \Delta z/(1+z) = $-0.1, -0.05 -=0, +0.05 and +0.10. The large red dots show final UVOT/GROND photo-$z$ measurements for real bursts where a spectroscopic redshift has been obtained (see Tab.~\ref{tab:photoz}). The green dot shows the photo-$z$ of the flat-spectrum radio quasar PKS 0537-286 derived in a similar manner ($z=3.10$; \citealp{2010A&A...509A..69B, 1978ApJ...226L..61W}).}
\label{accuracy}
\end{figure}

In the redshift range above $z>6.5$, two systematic effects bias the measurement (blue lines in Fig.~\ref{accuracy}). In contrast to the optical/UV filters, the NIR filters do not overlap due to emissivity of the NIR sky in certain wavelength ranges. In cases where the redshift signature is located in the filter gaps, it can not be measured precisely of course. For example, photometric measurements yield very similar SEDs for $z\sim7$ and $z\sim8$ afterglows with only upper ($z \lesssim 8.3$) and lower ($z \gtrsim 6.5$) limits on the photometric redshift, and there is hence a trend of over/under-estimating the redshift around $z\sim6.5-7$ and $z\sim7.5-8$, respectively. A $Y$-band filter centered at 1020~nm, as used in the analysis for GRB~090423 \citep{2009Natur.461.1254T, 2009Natur.461.1258S}, provides additional information in this redshift range. Furthermore, the continuum is only sparsely-probed at these redshifts. With three detections in $JHK_s$, different combinations of little dust, blue $\beta_o$, and a lower-$z$ as well as no dust, red $\beta_o$ and a higher redshift fit the input data equally well. While the returned best-fit photo-$z$ is based on the solution with lowest possible dust content, the input $A_V^{\rm host}$ values follow Fig.~\ref{synsample}. Hence, the average photo-$z$ of the sample consequently somewhat overpredicts the simulation input value, but with errors containing the input parameter space, of course. The redshift dependence of the average photo-$z$ (thick line in Fig.~\ref{accuracy}) is the net-result of both aforementioned effects.

The asymmetry of the blue-shaded error regions in the two lower panels of Fig.~\ref{accuracy} can be readily understood as the result of the upper systematic effects. The best-fit average photo-$z$ is based on the no-dust solution, but a family of solutions with somewhat higher dust values, which as a consequence also means lower redshifts, describe the simulated measurements equally well. This is represented by the asymmetric error regions extending to lower redshifts in the lower panels of Fig.~\ref{accuracy} and also illustrated in Fig.~\ref{contours}.

Furthermore, secondary and possibly tertiary solutions at lower redshifts and larger $A_V^{\rm host}$s of around $1-5\,\rm{mag}$ exist for afterglows where the continuum is constrained by only two filters, which includes all ultra high-$z$ events ($z\gtrsim 8.5$). Given a sufficiently bright afterglow, and hence accurate photometry, the low-redshift solutions can generally be ruled out at $>$99\% confidence under the assumption of a conventional dust attenuation law.

\begin{table}
\caption{Photo-$z$s of GRB afterglows compared against spectroscopic redshifts \label{tab:photoz}. Errors on spectroscopic measurements are reported when relevant ($>0.01$) and available.}
\begin{tabular}{cccc}
\hline
\noalign{\smallskip}
GRB & $z$ (spectroscopic) & $z_{\rm phot}$ &  References \\  
\hline
070802 & 2.4549 & $2.47^{+0.18}_{-0.15}$ & (1), (2) \\
071031 & 2.692 & $2.82^{+0.18}_{-0.15}$ & (3), (4), (5) \\
080129 & 4.349 & $4.18^{+0.14}_{-0.17}$ & (6) \\
080804 & 2.2045 & $2.11^{+0.09}_{-0.12}$ & (7), (8), (9), (10) \\
080913 & $6.70\pm0.03$ & $6.46^{+0.22}_{-0.19}$  & (11), (12), (13) \\
080928 & 1.692 & $1.73^{+0.08}_{-0.08}$ & (14), (15), (16) \\
081008 & 1.967 & $2.06^{+0.06}_{-0.09}$ & (17), (18), (19) \\
081028 & 3.038 & $3.12^{+0.15}_{-0.16}$ & (20), (21), (22) \\
081029 & 3.8479 & $3.77^{+0.14}_{-0.20}$ & (23), (24), (25), (26) \\
081121 & 2.512 & $2.59^{+0.12}_{-0.17}$ & (27), (28), (29) \\
090205 & 4.650 & $4.59^{+0.16}_{-0.12}$ & (30), (31), (32) \\
090313 & 3.375 & $3.20^{+0.25}_{-0.21}$ & (33), (34), (35) \\
090323 & 3.568 & $3.44^{+0.18}_{-0.16}$ & (36), (37) \\
090423 & $8.23^{+0.06}_{-0.07}$ & $8.0^{+0.4}_{-0.8}$ & (38), (39), (40) \\
090426 & 2.609 & $2.54^{+0.16}_{-0.17}$ & (41), (42), (43), (44) \\
090516 & 4.106 & $4.06^{+0.12}_{-0.15}$ & (45), (46) \\
090519 & 3.85 & $3.9^{+0.5}_{-0.8}$ & (47), (48) \\
090812 & 2.452 & $2.7^{+0.2}_{-0.3}$ & (49), (50), (51) \\
091029 & 2.752 & $2.6\pm0.2$ & (52), (53), (54) \\
100219A & 4.667 & $4.70\pm0.15$ & (55), (56) \\
\hline
\hline
\end{tabular}

\noindent{(1) \citet{2008ApJ...685..376K}, (2) \citet{2009ApJ...697.1725E}, (3) \citet{2007GCN..7028....1B}, (4) \citet{2009ApJ...697..758K}, (5) \citet{2007GCN..7023....1L}, (6) \citet{2009ApJ...693.1912G}, (7) \citet{2008GCN..8058....1T}, (8) \citet{2008GCN..8069....1K}, (9) \citet{2008GCN..8065....1C}, (10) \citet{2008GCN..8075....1K}, (11) \citet{2008GCN..8218....1R}, (12) \citet{2008GCN..8225....1F}, (13) \citet{2009ApJ...693.1610G}, (14) \citet{2008GCN..8298....1K}, (15) \citet{2008GCN..8296....1R}, (16) \citet{2008GCN..8301....1V}, (17) \citet{2008GCN..8346....1C}, (18) \citet{2008GCN..8350....1D}, (19) \citet{2010ApJ...711..870Y}, (20) \citet{2008GCN..8430....1S}, (21) \citet{2008GCN..8424....1C}, (22) \citet{2008GCN..8434....1B}, (23) \citet{2008GCN..8450....1H}, (24) \citet{2008GCN..8437....1C}, (25) \citet{2008GCN..8438....1D}, (26) \citet{2008GCN..8448....1C}, (27) \citet{2008GCN..8544....1O}, (28) \citet{2008GCN..8542....1B}, (29) \citet{2008GCN..8540....1L}, (30) \citet{2009GCN..8888....1K}, (31) \citet{2009GCN..8892....1F}, (32) \citet{2009GCN..8983....1U}, (33) \citet{2009GCN..8994....1C}, (34) \citet{2009GCN..9026....1U},  (35) \citet{2009GCN..9006....1S},
(36) \citet{2010arXiv1004.2900C}, (37) \citet{2009GCN..9215....1O}, (38) \citet{2009Natur.461.1254T}, (39) \citet{2009Natur.461.1258S}, (40) \citet{2009GCN..9265....1O}, (41) \citet{2009GCN..9268....1O}, (42) \citet{2009GCN..9264....1L}, (43) \citet{2009GCN..9269....1T}, (44) \citet{2009GCN..9382....1R}, (45) \citet{2009GCN..9383....1D}, (46) \citet{2009GCN..9408....1R}, (47) \citet{2009GCN..9409....1T}, (48) \citet{2009GCN..9771....1D}, (49) \citet{2009GCN..9773....1U}, (50) \citet{2009GCN..9774....1S}, (51) \citet{2009GCN.10108....1M}, (52) \citet{2009GCN.10098....1F}, (53) \citet{2009GCN.10100....1C}, (54) \citet{2010GCN.10439....1K}, (55) \citet{2010GCN.10443....1C}, (56) \citet{2010GCN.10445....1D}}
\end{table}

\subsection{The effect of photometric accuracy}
\label{photacc}

\begin{figure}
\includegraphics[width=\columnwidth]{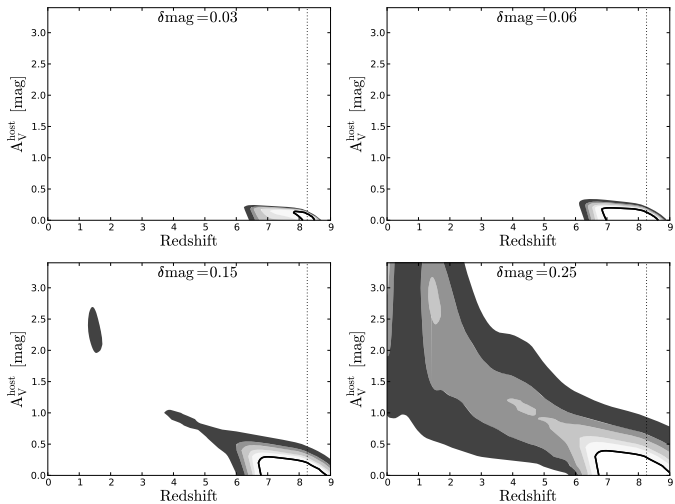}
\caption{Photometric redshift accuracy for GRB~090423 with different photometric errors. From top left to bottom right, the analysis was performed with the \gK~data of \citet{2009Natur.461.1254T} but with photometric errors of $\pm 0.03$~mag, $\pm 0.06$~mag, $\pm 0.15$~mag and $\pm 0.25$~mag respectively. The black line corresponds to the 1$\sigma$ contour, while the increasingly dark shaded areas correspond to the 90\%, 95.4\% (2$\sigma$), 99\%, 99.73\% (3$\sigma$) and 99.99\% confidence contours. The vertical dashed line denotes the spectroscopic redshift.}
\label{contours}
\end{figure}

The absolute accuracy in the photometric measurement is the basic quantity in the accuracy of photometric redshifts (see also \citealp[e.g.][]{bol00,2009ApJ...690.1236I}). Also the breaking of the dust-redshift degeneracy strongly depends on photometric accuracy in cases where the continuum spectrum is constrained by only few filters. Fig.~\ref{contours} shows the $z-A_V^{\rm host}$ contours of GRB~090423 ($z\sim8.2-8.3$, \citealp{2009Natur.461.1258S, 2009Natur.461.1254T}) for different fixed photometric errors. Here, we used \gK~measurements only (excluding the available $Y$-band imaging) for a direct comparison of the results. Two aspects are clearly apparent. Firstly, and already mentioned in Sec.~\ref{photozacc} and shown in Fig.~\ref{accuracy} is the asymmetric shape of the contours including regions of lower redshift with some dust contribution. Secondly, the large increase in the allowed parameter space when going from highly accurate ($\pm 0.03$~mag) to crude ($\pm 0.25$~mag) photometric measurements. In the latter case, photometry can no longer disentangle high-$z$ and large $A_V^{\rm host}$ solutions. The redshift is no longer constrained, and the 3$\sigma$ contour nearly contains the full redshift interval.

\subsection{Caveats in the photo-$z$ measurement}
\label{caveats}

The key uncertainty in the photometric redshift measurements is the attenuation of photons due to dust along the line of sight, which is generally assumed to be similar to local galaxies or to previously observed extinction laws. The dust extinction at high redshift, and in particular in extreme environments such as the circumburst medium of a GRB might be significantly different. The dust around the GRB might possibly be subject to destruction by the intense UV radiation of the afterglow or the GRB progenitor \citep{2000ApJ...537..796W, 2002ApJ...569..780D, 2007A&A...467..629D}, albeit a clear and highly significant observational signature of this process is still lacking \citep[e.g.][]{2009arXiv0912.2999P}. Although nearly all afterglow SEDs are well described with continuous extinction laws \citep[e.g.][]{2004ApJ...614..293S, 2004ApJ...608..846S, 2006ApJ...641..993K, 2007ApJ...661..787S, 2007MNRAS.377..273S, 2010MNRAS.401.2773S}, a very sharp break in the attenuation curve could possibly be misidentified as the signature of the Lyman breaks. This dust feature, however, must even be sharper than what was observed for GRB~070318 \citep{2009ASPC..414..277W}, so far the only case where such a break has been reported.

A second caveat is present in the case of sources located at regions with high Galactic foreground reddenings. The available foreground maps \citep{1998ApJ...500..525S} are limited to a spatial resolution of few arcminutes, and can hence be subject of uncertainties in the foreground correction of up to several tens of percent. For GRBs, typically at large Galactic latitudes with $E_{B-V}\lesssim0.1$ this effect is usually smaller than the absolute photometric accuracies and hence negligible. Extreme caution is however required when interpreting spectral breaks of objects close to the Galactic plane ($\mid b \mid \lesssim 5-10^\circ$) associated with regions of $E_{B-V}\gtrsim0.5$. A detailed analysis is crucial in these cases.

\begin{figure}
\includegraphics[width=\columnwidth]{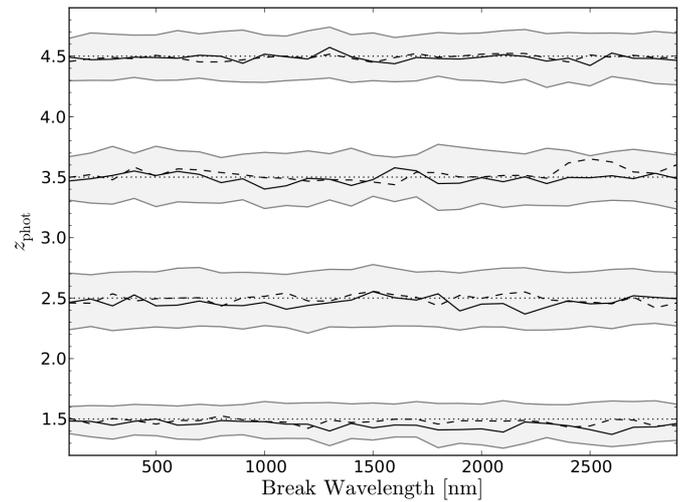}
\caption{Photometric redshift accuracy for four synthetic bursts ($z=1.5$, $z=2.5$, $z=3.5$ and  $z=4.5$) with a smoothly broken power law as continuum spectrum against the location of the break wavelength. Horizontal dashed lines mark the input redshift, and the best-fit photo-$z$ for an unbroken continuum, and black solid lines the average best fit photometric redshift when following Sec.~\ref{photozcode}. Grey shaded areas represent the typical 1$\sigma$ uncertainty, derived in similar manner as in Sec.~\ref{photozacc}.}
\label{ebl}
\end{figure}

Finally, there is the possibility that a break in the synchrotron spectrum, most likely the cooling break at $\nu_c$ is located in or evolves through the UV/optical/NIR range at the time of the observations. To quantify the effect of an evolving cooling frequency through the optical bands, the previously used continuum emission of a single power-law was replaced by a smoothly connected broken power-law \citep[e.g.][]{1999A&A...352L..26B, 2002ApJ...568..820G}. The difference between the two power-law slopes $\beta_1-\beta_2$ and the smoothness of the break were set to 0.5, and 1, respectively, following \citet{2002ApJ...568..820G} for a cooling break in the slow-cooling regime. Fig.~\ref{ebl} shows photometric redshifts obtained for four mock afterglows with a fixed flux at 1000~nm restframe of $60\,\mu Jy$, where the break wavelength evolves through the UV to the NIR. These synthetic afterglows are located at redshifts of $z=1.5$, $z=2.5$, $z=3.5$, and $z=4.5$, and hence span the range of the largest fraction of \textit{Swift} GRBs starting from the low redshift end of the photo-$z$ sensitivity. Their photo-$z$ with respect to break wavelength and compared to the standard single power law continuum spectrum is shown in Fig.~\ref{ebl}. The introduced curvature due to the cooling break is interpreted as an increased dust content, but can not resemble the strong breaks due to Lyman-blanketing. Hence, even in case of a broken power-law continuum spectrum, the photometric redshift can be considered reliable.

\subsection{Number statistics}

The total number of bursts where a photo-$z$ could in principle be derived with the presented method depends on a number of factors, primarily a UVOT or GROND detection and wavelength coverage of the redshift signature. Using the typical GROND and UVOT detection efficiency \citep{2009ApJ...690..163R, 2010AandAJochen} and the most recent GRB redshift distribution \citep{2009ApJS..185..526F}, and requesting a favorable declination for GROND observations and a Galactic latitude cut of $\mid b \mid > 10^\circ$, the presented method is applicable to around 15-30\% of all GRBs detected by \textit{Swift}/BAT, and 30-50\% of all GRBs with a detected optical afterglow. For many (70\%-90\%) of the latter sources spectroscopic observations will of course be feasible and return a much more accurate redshift measurement. Nevertheless, in few of all bursts ($\sim$1-5\%), which tend to include the rare LAT or ultra high-$z$ events \citep[e.g.][Cucchiara et al., in prep.]{2009A&A...498...89G} a photo-$z$ provides the best redshift measurement for GRBs and their afterglows.

\section{Application to individual bursts}
\label{appl}
To further demonstrate the concept of photometric redshifts for GRBs, a number of afterglows between July 2007, which is when GROND began systematic follow-up observations of all GRBs, until May 2010 were extracted from the UVOT and GROND archives. These afterglows must be located at regions of low Galactic foreground reddenings ($E_{B-V}<0.3$), detected by UVOT/GROND in at least three filters, and no spectroscopic redshift for the bursts is available. Finally, the SED of the afterglows must show a prominent break in the observed wavelength range to constrain the redshift to at least 90\% confidence. Five out of a total of $\sim$ 105 GROND observed bursts (i.e. around 5\%) meet the earlier constraints. Out of these five sources, four triggered \textit{Swift}/BAT and one triggered SuperAGILE. 

\subsection{Data reduction and cross calibration}

GROND data for the selected bursts shown in Tab.~\ref{tab:seds} were reduced in the standard manner using pyraf/IRAF \citep{1993ASPC...52..173T, 2008AIPC.1000..227Y}. The data in different filters are obtained simultaneously by hardware setup with the exception of a small difference in the mean photon arrival time between the \griz~and $JHK_s$ filters (see \citealp[e.g.][]{2009ApJ...697..758K}). At the average observing time of GROND after the GRB trigger of several hours, this difference is $\delta t/t_{\rm obs}\leq10^{-3}$, and hence negligible in the analysis. Absolute photometry for \griz~has been tied to the SDSS standard star network \citep{2002AJ....123.2121S} or nearby fields covered by the SDSS catalog DR7 \citep{2009ApJS..182..543A}. Photometry for $JHK_s$ has been derived against 2MASS field stars in all cases \citep{2006AJ....131.1163S}. The GROND measurements, and the respective calibration source are given in Tab.~\ref{tab:seds}.

\textit{Swift}/UVOT photometry has been obtained following \citet{2008MNRAS.383..627P}. As UVOT operates through filter cycles, the measurements had to be interpolated to a common epoch detailed in \citet{2010MNRAS.401.2773S}. Typically, the interpolation is performed over a short time interval, and the introduced uncertainties are much smaller than the individual measurement errors dominated by photon noise. UVOT photometry and the reference time of its observations are provided in Tab.~\ref{tab:uvotseds}.

\begin{table*}
\caption{GROND photometric measurements of GRB afterglows in the photo-$z$ sample \label{tab:seds}}
 
\begin{tabular}{ccccccccccc}
\hline
\noalign{\smallskip}
GRB & $\Delta t$ & $g^{\prime a}$ &  $r^\prime$ & $i^\prime$ & $z^\prime$ & $J$ & $H$ & $K_S$ & Calibration$^{(b)}$\\  
\hline
 & [h] & [mag$_{\rm AB}$] & [mag$_{\rm AB}$] & [mag$_{\rm AB}$] & [mag$_{\rm AB}$] & [mag$_{\rm Vega}$] & [mag$_{\rm Vega}$] & [mag$_{\rm Vega}$] & &  \\  
\hline
080825B & 7.20 & $22.45\pm0.07$ & $19.91\pm0.04$ & $18.62\pm0.04$ & $18.24\pm0.04$ & $16.83\pm0.04$ & $16.03\pm0.05$ &  $15.30\pm0.05$ & SA103-626/2M\\
080906  & 10.9 & $22.11\pm0.07$ & $21.72\pm0.05$ & $21.49\pm0.06$ & $21.37\pm0.07$ & $20.18\pm0.16$ & $19.55\pm0.23$ &  $18.75\pm0.32$ & SDSS/2M\\
081228  & 0.719 & $22.10\pm0.14$ & $20.79\pm0.04$ & $20.30\pm0.05$ & $20.12\pm0.08$ & $18.64\pm0.14$ & $17.63\pm0.14$ &  $16.87\pm0.15$ & SDSS/2M\\
081230  & 5.44 &  $21.62\pm0.04$ & $21.16\pm0.03$ & $20.98\pm0.04$ & $19.75\pm0.04$ & $19.62\pm0.05$ & $18.99\pm0.06$ &  $18.50\pm0.14$ & SA94-242/2M\\
090530  & 21.8 & $22.15\pm0.14$ & $22.08\pm0.10$ & $21.78\pm0.14$ & $21.64\pm0.18$ & $>20.33$ & $>19.50$ &  $>18.74$ & SDSS/2M\\
\hline
\end{tabular}

\noindent{$^{(a)}$ All magnitudes are observed magnitudes, i.e. uncorrected for the corresponding Galactic foreground reddening.\\
$^{(b)}$ Calibration source for the data. All magnitudes from SA standard stars and fields are taken from the primary Sloan standard star network \citep{2002AJ....123.2121S}. 2M and SDSS denote a calibration against stars from the 2MASS catalog \citep{2006AJ....131.1163S} in case of $J$, $H$, and $K_S$ measurements, and the SDSS DR7 \citep{2009ApJS..182..543A} for \griz~band data.}
\end{table*}

\begin{table*}
\caption{UVOT photometric data \label{tab:uvotseds}}

\begin{tabular}{cccccccc}
\hline
\noalign{\smallskip}
GRB & $\Delta t^{(a)}$ &  $uvw2^{(b)}$ & $uvm2$ & $uvw1$ & $u$ & $b$ & $v$ \\  
\hline
 & [h] & [mag] & [mag] & [mag] & [mag] & [mag] & [mag] \\  
\hline
080906 & 0.194 & $> 19.54$ & $> 19.01$ & $> 20.16$ & $19.05^{+0.40}_{-0.29}$ & $> 18.90$ & $18.93^{+0.31}_{-0.24}$ \\
081230 & 0.139 & $> 19.82$ & $> 19.62$ & $> 19.76$ & $19.31^{+0.30}_{-0.23}$ & $19.26^{+0.19}_{-0.16}$ & $19.61^{+0.73}_{-0.43}$ \\
090530 & 1.67 & $21.17^{+0.62}_{-0.39}$ & $19.74^{+0.25}_{-0.20}$ & $20.29^{+0.49}_{-0.34}$ & $19.51^{+0.06}_{-0.05}$ & $20.19^{+0.20}_{-0.17}$ & $20.36^{+0.48}_{-0.33}$ \\
\hline
\end{tabular}

\noindent{$^{(a)}$ UVOT magnitudes are the result of interpolation to the given reference time.\\
$^{(b)}$ All data are observed magnitudes, and in the UVOT system.}
\end{table*}

The reference time of UVOT measurements is generally much earlier than the ground-based observations, but the large overlap in the $bv$ and $g^\prime r^\prime$ filters (see Fig.~\ref{filters}) offers a straight-forward cross calibration. Using the synthetic photometry from the mock spectra in Sec.~\ref{photozgrb}, and including stellar templates, color terms are derived over a large range of photometric colors:

\begin{equation}
\label{bg}
b-g^\prime = 0.15(g^\prime -r^\prime)+0.03(g^\prime -r^\prime)^2\: \forall\, (g^\prime -r^\prime)  \in [-1, 2]
\end{equation}
\begin{equation}
v-r^\prime = 0.62(r^\prime -i^\prime)+0.10(r^\prime -i^\prime)^2\: \forall\, (r^\prime -i^\prime)  \in [-1, 2]
\end{equation}
\begin{equation}
g^\prime -b = -0.20(b-v)-0.05(b-v)^2\: \forall\, (b -v)  \in [-1, 2]
\end{equation}
\begin{equation}
r^\prime -v = -0.55(b-v)-0.04(b-v)^2\: \forall\, (b -v)  \in [-1, 2]
\end{equation}
where all UVOT magnitudes are in the AB system. The systematic error on the color terms is of order 2\% for the individual equations, and smaller than 5\% for all of them. Due to the large spectral overlap of the GROND and UVOT filters, the systematic uncertainties introduced through the cross calibration process are between 5 and 10~mmag even for SEDs with red colors of $g^\prime -r^\prime \sim 1$. This is much smaller than the individual measurement errors (see Tab.~\ref{tab:uvotseds}) and does not introduce additional uncertainty in the measurement.

If possible, the transformation equation with the largest spectral overlap (i.e., Eq.~\ref{bg}) is used to transform the UVOT measurements to the GROND epoch, where errors are propagated accordingly. This implicitly assumes that the overall spectral evolution of the optical transient associated with the burst is achromatic. While there is evidence from well-sampled optical afterglow light-curves of chromatic evolution simultaneous to strong variability \citep[e.g.][]{2009ApJ...693.1912G}, the associated color changes are generally moderate to absent even in complex panchromatic light-curves \citep[e.g.][]{2000A&A...359L..23M, 2003AJ....125.2291H, 2003A&A...404L...5C, 2004ApJ...606..381L, 2008Natur.455..183R, 2008ApJ...685..361U, 2008ApJ...672..449P, 2009ApJ...691..723B, 2009MNRAS.395..490O, 2009A&A...508..593K, 2009ApJ...693.1484C, 2009arXiv0912.2999P, 2010arXiv1003.3885M, 2007arXiv0712.2186K, 2010A&A...521A..53C}. Hence, even if a chromatic light-curve evolution is present, either due to the passage of the cooling break, or associated with late inner engine activity, it is very unlikely to mimic the strong breaks in the UV/optical/NIR SED due to absorption by neutral hydrogen used for the photo-$z$ measurement. In all cases, however, the hydrogen column of the DLA is unconstrained by the photometric measurement, and a DLA with $\log N_H \sim 21.5$, which is about the median value for afterglows where such a measurement was possible \citep[e.g.][]{2009ApJS..185..526F} is adopted in the SEDs (Figures \ref{080825B}, \ref{080906}, \ref{081228}, \ref{081230} and \ref{090530}). The properties of the five afterglows including their photometric redshifts are summarized in Tab.~\ref{tab:agprop}.

\begin{table}
\caption{Properties of the GRB afterglows in the photo-$z$ sample \label{tab:agprop}}
\begin{tabular}{cccc}
\hline
\noalign{\smallskip}
GRB & $z_{\rm phot}$ & $\beta$ &  $A_V^{\rm host}$ [mag] \\  
\hline
080825B & $4.31^{+0.14}_{-0.15}$ & $0.4^{+0.3}_{-0.2}$ & $0.20\pm0.15$ \\
080906 & $2.13^{+0.14}_{-0.20}$ & $0.5^{+0.1}_{-0.2}$ & $<0.3$ (2$\sigma$) \\
081228 & $3.44^{+0.15}_{-0.32}$ & $1.2^{+0.1}_{-0.5}$ & $0.03^{+0.17}_{-0.03}$ \\
081230 & $2.03^{+0.16}_{-0.14}$ & $0.4\pm0.2$ & $0.22^{+0.14}_{-0.08}$ \\
090530 & $1.28^{+0.16}_{-0.15}$ & $0.4\pm0.3$ & $0.15^{+0.15}_{-0.08}$ \\
\hline
\end{tabular}
\end{table}



\subsection{GRB 080825B}

\begin{figure}
\centering
\includegraphics[width=\columnwidth]{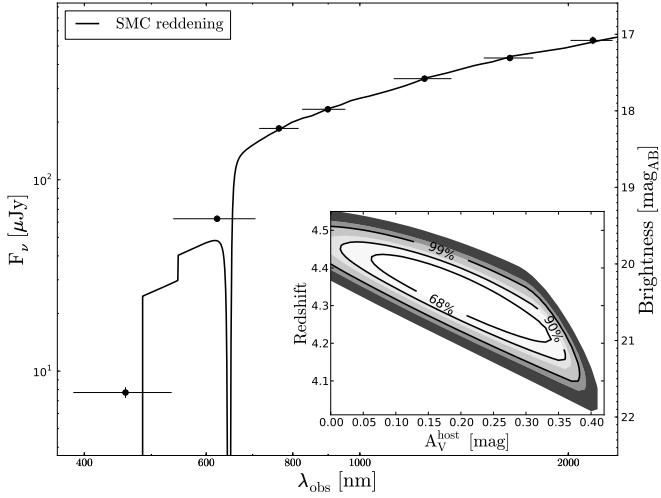}
\caption{Broad-band spectral energy distribution of the afterglow of GRB~080825B as observed with GROND. The foreground-corrected SED shows two prominent breaks corresponding to Lyman-limit absorption in the $g^\prime$ band, and Lyman-$\alpha$ in $r^\prime$. The corresponding redshift is $z_{\rm phot}=4.31^{+0.14}_{-0.15}$. The inset shows the confidence contours of the redshift solution versus intrinsic extinction, where the three-dimensional $z-\beta-A_V^{\rm host}$ parameter space has been collapsed onto a two-dimensional $z-A_V^{\rm host}$ grid. The increasingly dark shaded areas correspond to the 68.3\% (1$\sigma$), 90\%, 95.4\% (2$\sigma$), 99\%, 99.73\% (3$\sigma$) and 99.99\% confidence contours, where the 68.3\%,  90\% and 99\% contours are also marked with solid lines. The different filter bands are plotted with their effective wavelength, and the horizontal error bars represent their FWHM, i.e. from 50\% to 50\% of maximum transmission.}
\label{080825B}
\end{figure}

GRB~080825B was detected by AGILE \citep{2008GCN..8133....1E} and Konus-Wind \citep{2008GCN..8142....1G}, and the optical/X-ray afterglow was rapidly identified by \citet{2008GCN..8135....1T} and \citet{2008GCN..8140....1P}. The \gK~GROND SED obtained at a midtime of 7.2~h after the trigger (Fig.~\ref{080825B}) shows the clear presence of two strong breaks between $g^\prime$ and $r^\prime$ and $r^\prime$ and $i^\prime$, respectively (see Fig.~\ref{080825B}). In addition there is evidence for curvature red-wards of 650~nm which is well described with a SMC-type reddening. The best-fit ($\chi^2=1.68$ for 3 d.o.f) photometric redshift is $z_{\rm phot}=4.31^{+0.14}_{-0.15}$ with a host extinction of $A_V^{\rm host}=0.20\pm0.15$ and a spectral index of $\beta=0.4^{+0.3}_{-0.2}$, consistent with the redshift limits from \citet{2008GCN..8137....1D}. At this redshift, and using the $\gamma$-ray properties from \citet{2008GCN..8142....1G}, the isotropic equivalent energy release of the GRB in the rest-frame 1~keV to 10~MeV energy range is $\approx3.7\times10^{53}$~erg with a rest-frame peak energy of $\sim380$~keV. 

\subsection{GRB 080906}

\begin{figure}
\centering
\includegraphics[width=\columnwidth]{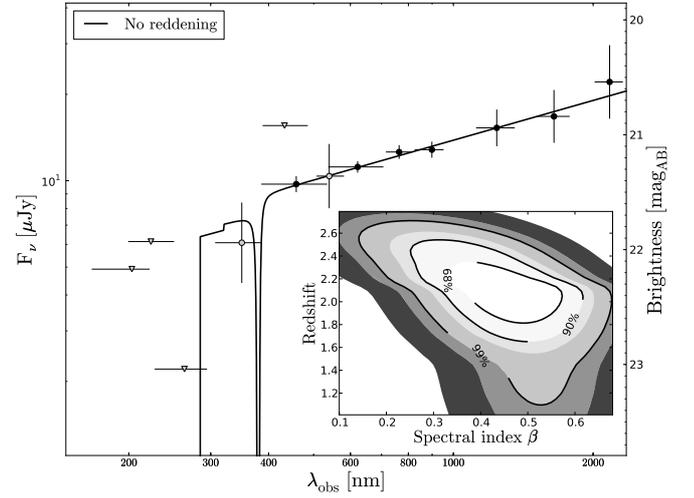}
\caption{Foreground corrected broad-band spectral energy distribution of the afterglow of GRB~080906 with UVOT (open circles) and GROND (filled circles). Upper limits are shown by downward triangles. The SED shows a strong break due to the Lyman-limit being located in the $uvw1$ band. The photometric redshift is $z_{\rm phot}=2.13^{+0.14}_{-0.20}$. The inset shows the confidence contours of the redshift solution versus spectral index, where the three-dimensional $z-\beta-A_V^{\rm host}$ parameter space has been collapsed onto a two-dimensional $z-\beta$ grid. Lines and shadings are the same as in Fig.~\ref{080825B}.}
\label{080906}
\end{figure}

GRB~080906 \citep{2008GCN..8186....1V} triggered \textit{Swift}, and its optical afterglow was detected by UVOT 82~s \citep{2008GCN..8198....1H} and with GROND 10.7~h \citep{2008GCN..8194....1A} after the burst. The combined SED extends from $u$ to the $K_s$ band and is shown in Fig.~\ref{080906}. The photometric redshift is driven by the $uvw1$ non-detection and is $z_{\rm phot}=2.13^{+0.14}_{-0.20}$ with negligible intrinsic host extinction and a spectral index of $\beta = 0.5^{+0.1}_{-0.2}$. Previous claims from \citet{2008GCN..8212....1H} are consistent with this redshift.

\subsection{GRB 081228}

\begin{figure}
\centering
\includegraphics[width=\columnwidth]{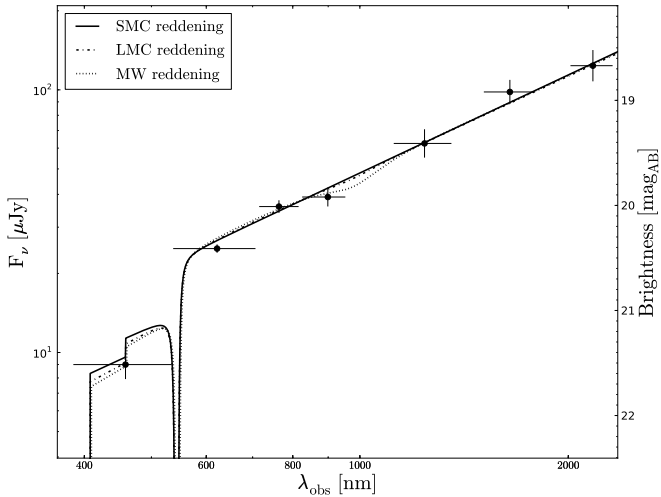}
\caption{Foreground corrected broad-band spectral energy distribution of the afterglow of GRB~081228 as observed with GROND. The SED shows a prominent break corresponding to Lyman-$\alpha$ in the $g^\prime$ band. The resulting redshift is $z_{\rm phot}=3.44^{+0.15}_{-0.32}$.}
\label{081228}
\end{figure}

\textit{Swift}/BAT triggered on 
GRB~081228 \citep{2008GCN..8742....1P}, and GROND detected the optical afterglow \citep{2008GCN..8745....1A}, while UVOT observations only yield upper limits \citep{2008GCN..8747....1L}. The SED in the GROND filters is dominated by a red $g^\prime-r^\prime$ color of $\sim1.3$~mag \citep{2008GCN..8752....1A}, which yields a photometric redshift of $z_{\rm phot}=3.44^{+0.15}_{-0.32}$. There is no strong evidence of excess absorption with a best fit spectral index $\beta=1.24^{+0.11}_{-0.46}$ with $A_V^{\rm host}=0.03^{+0.17}_{-0.03}$ assuming a SMC type reddening law ($\chi^2 = 2.34$ for 3 d.o.f). The SED is slightly better fit with a 2175\AA~feature ($z_{\rm phot}=3.49^{+0.13}_{-0.23}$ and $\chi^2 = 1.58$ for MW-like reddening for 3 d.o.f, and $z_{\rm phot}=3.45^{+0.14}_{-0.29}$ and $\chi^2 = 2.14$ for LMC-like reddening for 3 d.o.f, see Fig.~\ref{081228}), which, given the small improvement in $\chi^2$, is only significant at the 0.9 and 0.5$\sigma$ level, respectively.

\subsection{GRB 081230}

\begin{figure}
\centering
\includegraphics[width=\columnwidth]{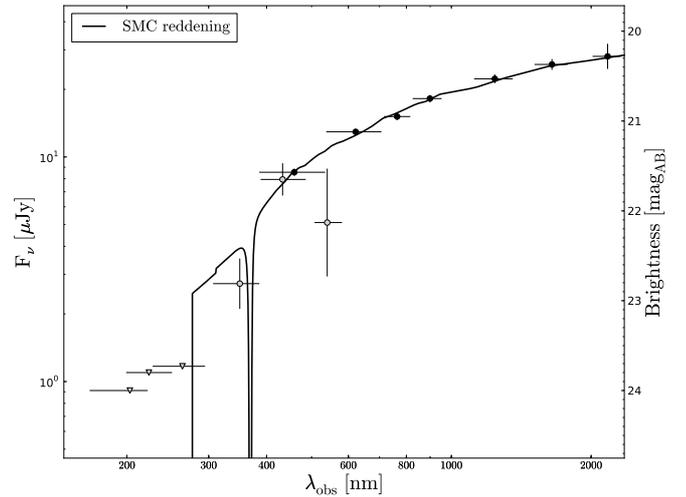}
\caption{Foreground corrected broad-band spectral energy distribution of the afterglow of GRB~081230 as observed with UVOT (open circles) and GROND (filled circles). Upper limits are shown as downward triangles The SED shows a prominent break corresponding to Lyman-limit absorption between $uvw1$ and $u$ bands. The resulting redshift is $z_{\rm phot}=2.03^{+0.16}_{-0.14}$.}
\label{081230}
\end{figure}

The UV/optical/NIR afterglow of the \textit{Swift} GRB~081230 \citep{2008GCN..8753....1L} was imaged by UVOT starting 131~s \citep{2008GCN..8757....1O} and GROND 4.2~h after the trigger \citep{2009GCN..8760....1A}. The combined SED (Fig.~\ref{081230}) is broad, ranging from the $u$ to $K_s$ filter, whereas the afterglow is undetected in the UV bands. The implied break between the $uvw1$ and $u$ bands yields a photometric redshift of $z_{\rm phot}=2.03^{+0.16}_{-0.14}$. There is mild curvature in the SED which is well fit ($\chi^2 = 8.2$ for 9~d.o.f) with a moderate amount of SMC-type reddening with an $A_V^{\rm host}=0.22^{+0.14}_{-0.08}$ and a spectral index $\beta=0.4^{+0.2}_{-0.2}$. LMC and MW-like reddening models are strongly ruled out due to the absence of a 2175\AA~dust feature and are not shown in Fig.~\ref{081230}.

\subsection{GRB 090530}

\begin{figure}
\centering
\includegraphics[width=\columnwidth]{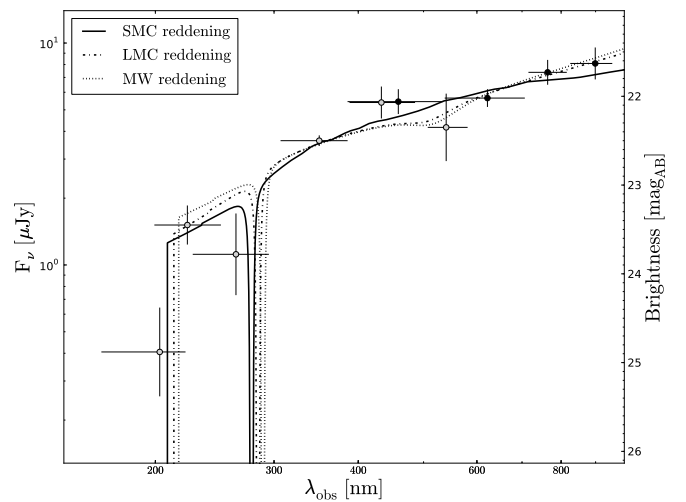}
\caption{Foreground corrected broad-band spectral energy distribution of the afterglow of GRB~090530 as observed with UVOT (open circles) and GROND (filled circles). The SED exhibits a break indicative of Lyman-limit absorption between the $uvw2$ and $uvm2$ bands. The corresponding redshift is $z_{\rm phot}=1.28^{+0.16}_{-0.15}$.}
\label{090530}
\end{figure}

GRB~090530 was detected by \textit{Swift} \citep{2009GCN..9438....1C}, and its afterglow was observed with space- and a number of ground-based telescopes \citep[e.g.][]{2009GCN..9441....1N, 2009GCN..9439....1F}. UVOT \citep{2009GCN..9450....1S} and GROND \citep{2009GCN..9458....1R} observations started 2.5~min and 21.3~h after the bursts. The combined UVOT/GROND SED is shown in Fig.~\ref{090530} and contains all filters except the NIR, which given the late GROND observations only yield non-constraining upper limits. The data are acceptably fit ($\chi^2 = 7.4$ for 6 d.o.f.) with a marginally ($A_V^{\rm host}=0.15^{+0.15}_{-0.08}$) SMC-type reddened power law of spectral index $\beta = 0.4\pm 0.3$. LMC and MW extinction models provide slightly worse fits to the data, but within the errors comparable redshifts, extinctions, and spectral indices. The dominating spectral feature is a break between the two bluest UVOT filters, which implies a photometric redshift of $z_{\rm phot}=1.28^{+0.15}_{-0.17}$, $z_{\rm phot}=1.33^{+0.17}_{-0.16}$, $z_{\rm phot}=1.37^{+0.16}_{-0.15}$ for SMC, LMC and MW extinction laws, respectively. Lower-redshift ($z<1$) solutions are allowed at the $2\sigma$ level.

\section{Conclusions}
 
GRB afterglow photometry in multiple bands offers a viable and robust method to derive the distance scale to the burst to reasonable accuracy. In cases where the photometric observations cover the wavelength range blue- and redwards of the redshift tracers such as the Lyman-limit in case of low $z\lesssim 5$, or Lyman-$\alpha$ for high ($z \gtrsim 3$) redshift GRBs, photo-$z$s can be obtained in principle for redshifts from $z\sim1$ out to $z\sim12$ for bright events with state-of-the-art follow up \citep[e.g.][]{2009MNRAS.395..490O, 2009ApJ...691..723B, 2009Natur.461.1254T, 2009Natur.461.1258S}. The accuracy of the photometric redshift determination is a function of redshift, wavelength coverage and quality of the photometric measurement, and can reach relative errors as good as $\eta\sim2-3$\% for observations as routinely performed by systematic follow-up with \textit{Swift}/UVOT and GROND. 

The photo-$z$ method presented in this work is based on standard template fitting, and derives the best-fit solution of redshift, reddening and spectral index of the afterglow against the obtained data simultaneously. In this way, redshifts for five GRBs between $z=1.2$ and $z=4.3$ with relative errors $\eta < 0.1$ and without spectroscopic observations are derived using photometric UV/optical/NIR data from UVOT and GROND.

Due to the simple synchrotron continuum, and the absence of strong reddening observed for a good fraction of all afterglows (\citealp[e.g.][]{2007arXiv0712.2186K, 2010AandAJochen}), and in particular the high-redshift ones \citep[e.g.][]{2007AJ....133.1187K, 2007ApJ...669....1R, 2009ApJ...693.1610G, 2009Natur.461.1254T, 2009Natur.461.1258S, 2010arXiv1002.4101Z}, two filters can constrain the spectrum such that a strong break due to Ly-$\alpha$ can reliably be identified in a third one. There is, however, a strong observational bias against highly-redshifted, highly-reddened afterglows, and a first moderately extinguished, $z\sim 5$ afterglow has only been detected recently \citep{2009arXiv0912.2999P}. Follow-up observations in the standard NIR broad-band filters can yield robust photo-$z$ measurements for bright events up to distance scales where Lyman-$\alpha$ is redshifted into the $H$ band. Due to the uncertainty in the dust extinction, around 10\% or 30\% of the well-detected high-z events beyond $z=9$ are expected to be deviating from the true redshift with more than 15\% or 10\%, respectively. It must be stressed, again, that these results are obtained from simulated afterglows with pre-defined reddening properties and brightnesses in a way that they are amenable to detailed photometric studies.

The strong breaks, their characteristic prominence and the well-defined continuum emission also provide a unique redshift solution under the assumption of a conventional dust attenuation law, and hence prevent catastrophic errors due to a degeneracy in spectral features such as a confusion between the Lyman-limit and the 4000~\AA~break for galaxies \citep[e.g.][]{1997AJ....113....1S}.

Albeit with much larger uncertainties than spectroscopic redshifts, the accuracy of afterglow photo-$z$s is adequate for further studies about the afterglow reddening along the line of sight \citep[e.g.,][]{2009A&A...498...89G}, the soft X-ray absorption \citep[e.g.,][]{2008A&A...491L..29R}, the GRBs energy budget \citep[e.g.][]{2009A&A...498...89G, 2009Sci...323.1688A, 2009A&A...508..173A}, its emission mechanisms \citep[e.g.][]{2009Sci...323.1688A, 2009MNRAS.400L..75K, 2009ApJ...700L..65Z} and the role of GRBs as probes for quantum gravity \citep[e.g.][]{2009PhRvD..80h4017A, 2009PhLB..674...83E}, the extragalactic background light \citep[e.g.][]{2009MNRAS.399.1694G} or demographic studies of GRBs \citep[][]{2006A&A...447..897J, 2009ApJS..185..526F}. Furthermore, photometric observations reach deeper flux limits than spectroscopy in particular in the NIR, which is of primary importance for redshifts $z>7$. In case of these ultra-high redshift events, an identification through photometry coupled with follow-up of their hosts with the upcoming generation of telescopes and instruments could reveal their hosts, and hence possibly a galaxy in the process of its first star formation.

\begin{acknowledgements}
We thank the referee for a helpful report as well as A. Cucchiara and A. J. Levan for very constructive comments on the manuscript. TK acknowledges support by the DFG cluster of excellence 'Origin and Structure of the Universe'. Part of the funding for GROND (both hardware as well as personnel) was generously granted from the Leibniz-Prize to Prof. G. Hasinger (DFG grant HA 1850/28-1). SMB acknowledges support of a European Union Marie Curie European Reintegration Grant within the 7th Program under contract number PERG04-GA-2008-239176. ARo and SK acknowledge support by DFG grant Kl 766/11-3. This research has made use of the data obtained from the High Energy Astrophysics Science Archive Research Center (HEASARC).

\end{acknowledgements}

\hyphenation{Post-Script Sprin-ger}


\end{document}